\documentclass[11pt, oneside]{article}   	
\usepackage{geometry}                		
\usepackage{graphicx}				
\usepackage{amssymb}
\usepackage{caption}
\usepackage{subcaption}
\usepackage{authblk}
\usepackage{booktabs}
\usepackage{tabularx}
\usepackage{rotating}
\usepackage{amsmath}
\usepackage{amsthm, amsfonts}
\usepackage{url}

\title{Generative AI Takes a Statistics Exam: A Comparison of Performance between ChatGPT3.5, ChatGPT4, and ChatGPT4o-mini}

\title{Maximizing Diver Score by Examining Discrepancies in Diver Competency and Judges' Marks}
\author{Monnie McGee\thanks{mmcgee@smu.edu} \& Bivin Sadler\thanks{bsadler@smu.edu} }
\affil{Department of Statistics and Data Science\\ Southern Methodist University\\ 6425 Boaz Lane, Dallas, TX, 75205}
\date{January 15, 2025}							

\begin{document}

\maketitle

\begin{abstract}

Many believe that use of generative AI as a private tutor has the potential to shrink access and achievement gaps between students and schools with abundant resources versus those with fewer resources. Shrinking the gap is possible only if paid and free versions of the platforms perform with the same accuracy. In this experiment, we investigate the performance of GPT versions 3.5, 4.0, and 4o-mini on the same 16-question statistics exam given to a class of first-year graduate students. While we do not advocate using any generative AI platform to complete an exam, the use of exam questions allows us to explore aspects of ChatGPT's responses to typical questions that students might encounter in a statistics course. Results on accuracy indicate that GPT 3.5 would fail the exam, GPT4 would perform well, and GPT4o-mini would perform somewhere in between. While we acknowledge the existence of other Generative AI/LLMs, our discussion concerns only ChatGPT because it is the most widely used platform on college campuses at this time. We further investigate differences among the AI platforms in the answers for each problem using methods developed for text analytics, such as reading level evaluation and topic modeling. Results indicate that GPT3.5 and 4o-mini have characteristics that are more similar than either of them have with GPT4.

 \end{abstract}

  
\section{Introduction}%
\label{sec:intro}

ChatGPT \cite{gpt3} publicly premiered in November of 2022 and upended the world of education. As with the calculator \cite{wapo}, laptop \cite{laptops}, and smartphone \cite{smartphones}, educators quickly took sides between ``ban AI from the classroom'' to ``free use of AI in the classroom'' and even ``champion the use of generative AI in the classroom.'' Many believe ChatGPT and other generative AI platforms will revolutionize education, providing a personalized tutor for every student \cite{khanmigo_article, KhanAcademy2024}. Access to a personalized tutor, even a virtual one, has the potential to diminish the gap between the educational backgrounds of students \cite{ellis23}. However, for this promise to be realized, students will need wireless access and a capable laptop computer. Students from difficult financial circumstances or infrastructure deserts will have more difficulty utilizing laptops, smartphones, and other items that are critical to their education \cite{tough19}. Some students and educators now feel that  a subscription to a  generative AI platform is critical to education, particularly for students who need personalized extra help \cite{kovari2024}. While use of generative AI is prohibited in some college classrooms, this is hard to police, and students likely will use it regardless of prohibitions \cite{terry}. When it is allowed, some students will be able to pay for more accurate and comprehensive versions, while others will need to use the free version. A significant difference in performance between the free and paid version of ChatGPT would ironically widen the digital and educational gaps even further. 

In 2023, OpenAI released the more powerful and quite impressive GPT4 \cite{openai2024gpt4}, but at a cost of \$20 per month. In May of 2024, OpenAI released ChatGPT4o \cite{gpt4o}, which is purported to have enhanced accuracy and precision, improved response time, greater ability to handle complex queries, new features, and better adaptability to languages other than English than previous versions of ChatGPT \cite{aiproteam}. Users on the ``free tier'', which is Open AI's designation for users who do not pay a monthly access fee, can use GPT4o as the default, but the number of messages is limited and can be further limited during peak demand. Free-tier users also have more limited access than subscribers (``Plus'' members) to do data analysis, file uploads, and image comprehension. In mid-summer of 2024, OpenAI introduced GPT4o-mini, which like GPT3.5, is available on the free tier \cite{lacey2024}.  Less than a month later, GPT3.5 was retired and replaced by GPT4o-mini \cite{lacey2024}. There is some ability for free tier users to use the most powerful OpenAI version, GPT4o; otherwise, free tier users can access GPT4o-mini within certain restrictions, which are subject to usage demand. In this paper, when referring to a specific version of ChatGPT, we will use the prefix ``GPT'' followed by the version number ({\it e.g.} GPT3.5). To refer to all Open AI's platforms, we will use the term ``ChatGPT''. 

A large body of literature is emerging about the benefits and pitfalls of generative AI in the classroom, for all levels of students. Researchers acknowledge the potential for misuse of this powerful technology and the problems of equitable access to the latest technology \cite{lawrence2024}. They also acknowledge that plagiarism is problematic and bias is present in LLMs \cite{aamir2024}. Indeed, these are important issues to be addressed, particularly as the use of generative AI platforms becomes ubiquitous.  At the same time, most of the discussion is about generative AI writ large, and not about different platforms, different tiers of access, or different disciplines. All of these must be considered if generative AI is to be used regularly as an educational tool. In this paper, we investigate ``equitable access'' in terms of differences between free-tier and plus-tier versions of ChatGPT. Specifically, we seek to quantify the differences in accuracy, reading level, and topics modeled in text output to questions on statistics exams of various types and at various levels using text analysis methodology. Here, we compare GPT3.5, GPT4, and GPT4o-mini using text analytics for topic modeling. We hypothesize that free generative AI platforms perform worse than paid generative AI platforms, and thus those that use them will have inferior and sometimes misleading results.  In Section \ref{sec:LitRev} we explain current literature on comparisons in accuracy of output among OpenAI's  various platforms and discuss previous usage of text analytics for generative AI. In Section \ref{sec:bg} we give the results of previous work comparing output between GPT3.5 and GPT4. Next, we explain our methods in Section \ref{sec:methods}, including the exams used for the first phase of the project: determining accuracy of the output from generative AI platforms. In Section \ref{sec:results} we give results of accuracy assessments on standardized exams, plus results of topic modeling for a graduate exam in statistics, given to first-year statistics students in October of 2022, before the widespread availability of generative AI and large language models. We summarize our work and posit directions for future research in Section \ref{sec:disc}.

\section{Previous Research on Differences in Performance of OpenAI Platforms}\label{sec:LitRev}

 The difference in performance between GPT3.5 and GPT4 has been examined quite extensively, particularly with respect to standardized exams. In some sense, the comparison to GPT3.5 is moot because it has been retired from general use; however, studies addressing its accuracy are worth noting if only as a baseline for performance. GPT3.5 was shown to perform poorly on the bar exam \cite{chat35bar}; however, GPT4 aced the exam \cite{chat4bar}. Similarly, GPT4 performed at a high level for the GRE, USA Biology Olympiad Semifinal Exam, and the Chartered Financial Analyst (CFA) exam \cite{exams}. Comparisons of performance have also been made for orthopaedic assessment examinations \cite{massey}, ophthalmology oral boards \cite{taloni}, and other medical assessments, with GPT3.5 performing worse or much worse than GPT4. A comparison between the two versions has been made on USMLE soft skill assessments, as well \cite{brin}. \cite{joshi}, \cite{huang23}, and \cite{Hidayatullah_2024} examined GPT3.5's ability to answer questions in computer science, foreign language, and English writing classrooms, respectively, with mixed results. 

To our knowledge, no one has quantified the differences among the output of GPT3.5, GPT4, and now GPT4o-mini for questions in statistics and data science using text analytics, such as reading level analysis, word counts, and topic modeling, as we do in this work. The accuracy of output from GPT3.5 has been examined in the context of homework problems from computer science \cite{joshi}. The authors specifically addressed the strengths and weaknesses of GPT3.5 when answering several types of questions, including true/false, multiple choice, multiple selection, short answer, long answer, design-based, and coding questions. Their investigation used questions collected from four different computer science courses at well-established universities, as well as questions from the Graduate Aptitude Test in Engineering (GATE).  GATE is a national examination given in India to examine undergraduate students' understanding in various disciplines \cite{gate}. The authors found that GPT3.5 had 92.8\% accuracy for coding questions and only 33.4\% accuracy for answering questions about database management systems. As for type of question, GPT3.5 answered true\/false questions with 76\% accuracy. Accuracy for other types of questions ranged from 39.5\% to 58.3\% \cite{joshi}. The authors examined various subjects and questions using only GPT3.5 and only for computer science questions. In addition, all of their quantitative assessments were based on accuracy of the answers. The conclusion of the paper is that GPT3.5 does not make a good tutor for students when studying for various assessments because its performance is extremely variable.

Interestingly, accuracy improved when the users added context to the question within the prompt, or reminded GPT3.5 of the context as different questions were asked  \cite{joshi}. For example, GPT3.5 achieved the maximum accuracy of 92.8\% when it was told to pretend it was ``a computer science undergraduate student preparing for technical interviews''.  The finding that context changes accuracy has been mentioned in other papers \cite{callanan2023, ball2024can, yao2024}. Not only does context affect accuracy, but even small changes in prompt wording can affect accuracy  \cite{ball2024can}. In the current paper, we  did not change any wording of the question in order to mimic how we thought students might use ChatGPT to get help on homework questions; therefore, we cannot comment on how providing context or changing question wording will affect our results on accuracy, reading level, word count, or topic modeling.  We reserve an investigation of the effect of ``prompt engineering'', including different frameworks of prompt engineering \cite{yao2024}, on text analytics from output for various generative AI platforms for future research.

\subsection{Topic Modeling for ChatGPT}

Research with respect to ChatGPT and topic modeling can be classified into the same three main groups: 1) topic modeling with ChatGPT, 2) topic modeling of public responses to ChatGPT (eg. Tweets) and 3) topic modeling of responses directly from ChatGPT.  

With respect to the first group, a study from \cite{Alharbi_2024} proposed a new methodology of topic modeling that paired Linear Discriminant Analysis (LDA) with ChatGPT to interpret topics of social media posts from a relatively small amount of text.  Another study, \cite{Rijcken_2023} assessed ChatGPT's ability to model topics through a three-stage process.  The first step was to model the topics of a corpus of texts using established topic modeling techniques.  The second step involved having both the domain expert and ChatGPT interpret these topics.  The third step involved comparing the human and ChatGPT interpretations in which the study found about half of the ChatGPT responses to be useful.  With respect to topic modeling of public responses to ChatGPT, \cite{Koonchanok_2024} studied the topics appearing in data from X.  The researchers found that from Dec 2022 to June 2023, the most frequently discussed topics were ``Education, Bard, Search Engines, OpenAI, Marketing, and Cybersecurity, but the ranking varies by month'', and the tweets tended to be correlated with the author's occupation.  

Studies focused on the topic modeling of the raw responses, narratives or replies from ChatGPT (or other generative AI) are beginning to populate the literature. Most of these deal with the impact of prompt engineering on output. \cite{mervaala2024} examined the effect of context building on keywords, number of words, and topics generated. They entered questions into ChatGPT with no modification (zero-shot) or various levels of context added. They conclude that the amount of context has an effect on keywords and topics, and that adding context reduces the variability in the text analytics. \cite{callanan2023} found that changing one word in a prompt could change the response in important ways. No research was found in which text analytics was applied to output from questions of the type that students might ask for additional help in the a given subject. Therefore, the current manuscript is unique in this respect, as we analyze the output from three versions of ChatGPT in response to questions asked from a statistics exam. The aim of our study is to determine whether any ChatGPT platform gave reasonable accuracy with respect to statistical methods and what differences exist among the accuracy and topics from the output of ChatGPT platforms.

\section{Accuracy for Statistics Exams}\label{sec:bg}
Previous work involved the comparison of the accuracy of output from GPT3.5 and GPT4 on nationally normed statistics exams and an exam given in a graduate course in statistical methods \cite{mcgee2024}. The standardized exams assessed were the Comprehensive Assessment of Outcomes in Statistics Exam (CAOS)\cite{caos}, the Advanced Placement (AP) Exam, and the Arkansas Council of Teachers of Mathematics Exam (ACTM) \cite{ACTMExam}. In addition the output from questions on a ``homemade'' exam given to seven first-year Ph.D. students in statistics and biostatistics in the fall of 2022 was examined. The study had two purposes: to determine whether there is a substantive difference in the correctness of the output from the paid and free versions of generative AI platforms (which were GPT3.5 and GPT4, respectively, at the time), and to determine whether any differences were shaped by the type of question asked. The three standardized exams are multiple choice exams, but some questions involve the interpretation of tables and charts. The graduate exam contained a mix of short answer and multiple choice questions. All exams were entered, question by question, into both ChatGPT3.5 and ChatGPT4; thus every question was answered using both platforms. The percentage correct and the total number of questions for each exam is given in Table \ref{tab:previous}.

\begin{table}[h]
  \caption{Percentage correct on standardized exams for GPT3.5 and GPT4. The second column contains the number of questions on each exam.}%
  \label{tab:previous}
  \centering
    \begin{tabular}{cccc}
\toprule
Exam & Questions & GPT3.5 Score & GPT4 Score \\
\midrule
ACTM & 25 & 64\% & 100\%\\
AP2011 & 22 & 50\% & 81\% \\
CAOS & 40 & 48\% & 70\% \\
First-Year Exam & 16 & 41 & 82\\
\bottomrule
\end{tabular}
\end{table}

The results in the table are listed in reverse order of difficulty of the exam. The ACTM exam is focused toward students in high school. A good score on the AP2011 exam determines whether a student is ready to take a college-level statistics course beyond the introductory statistics course. The CAOS exam is meant for college students who have completed an introductory statistics course, and the graduate exam was given in a first-year course for Ph.D. students in statistics. The scores for both generative AI platforms decrease with increasing difficulty of the exam, except that GPT4 performed better on the graduate exam than it did on the CAOS exam. Interestingly, GPT3.5 does not pass any of the exams at the traditional passing level of 70\%, and GPT4 passes all of the exams. It is clear from the scores that GPT4, the most advanced generative AI platform at the time, is not always accurate.

\cite{mcgee2024} also performed a McNemar's test to examine the discrepancy between concordant and discordant pairs. Concordant pairs in this context are questions for which both platforms were incorrect (11 questions) or both were correct (41 questions). Discordant pairs were those for which GPT4 was correct and GPT3.5 was incorrect (35 questions) or vice versa (6 questions). The null hypothesis for McNemar's test is that discordant pairs are equally likely to occur in either direction. For questions from the three nationally normed exams tested, McNemar's $\chi^2_1$ statistic had a value of $19.9$, which results in a p-value of $.000012$. Therefore, there is a 12 in 10 million chance of seeing a test statistic this smaller or smaller if the null hypothesis is true. There is extremely strong evidence against the null hypothesis in this case, indicating that one of the exams is more likely to be correct.

\cite{mcgee2024} also showed that GPT3.5 had much worse performance if the question involved interpretation of an image, for example, a boxplot or a histogram. Of the 30 questions across all three exams requiring interpretation of an image, GPT3.5 was incorrect on all 30, while GPT4 answered 20 of  the 30 correctly. The fact that GPT3.5 is a text-only generative AI platform made up almost all of its inaccuracy. In fact, an ordinal logistic regression analysis showed that GPT4 is on average 70\% more likely to provide a higher quality response than GPT3.5 if the question contains an image. For the response variable of the ordinal logistic regression analysis was 0 (GPT3.5 has the better answer) to 4 (GPT4 has a much better answer than GPT3.5). A value of 1 indicated that the answer on the question was of the same quality for both platforms, and a value of 2 indicated that GPT4's answer was marginally better than GPT3.5's answer, while a value of 3 indicated that the answer for GPT4 was somewhat better than the answer for GPT3.5. The quality of the answers was determined first by independent examination by the authors and then by consensus discussion for any questions on which there was disagreement.

According to previous work, GPT4, which requires a subscription of \$20 per month, gives more accurate and better quality answers than GPT3.5. If generative AI is to be used in statistics courses as a tutor or in some other fashion, the ability of a student to pay for a subscription will determine the quality of the instruction. This puts students who cannot or will not pay for a subscription to GPT4 at a disadvantage. However, the landscape has changed now that GPT3.5 has been deprecated, and GPT4o-mini has taken its place. In the rest of this paper, we examine differences in reading level and relevance for the graduate exam, with 16 questions, among GPT3.5 (included as a baseline), GPT4, and GPT4o-mini. Our analysis includes topic modeling, and shows differences among the topics present among the platforms in the responses to the exams. The previous analysis of nationally normed exams did not involve an analysis of the text generated as answers to questions in any formal way beyond the overall correctness with respect to statistical accuracy; therefore, text generative by ChatGPT that explained its responses for the AP, ACTM, and CAOS exams is not analyzed in this study. 

\section{Methods}\label{sec:methods}

Exam questions from a midterm exam in a first-year graduate statistical methods course were loaded one-by-one into GPT3.5, GPT4, and GPT4o-mini. This was done in a ``zero-shot'' framework \cite{yao2024}, meaning that the question was entered verbatim from the exam with no context or modifications for clarification. Seven students had taken the exam in October of 2022 before generative AI platforms were widely available. Answers were assessed and grades were calculated from the students' answers and from the three versions of GPT.  Text of the answers generated by GPT3.5, GPT4, and GPT4o-mini were imported into R (version 4.3.2) packages \texttt{quanteda} and \texttt{readtext} \cite{rcore,quanteda,readtext}, to examine differences in the text of the answers. Stopwords were removed from the text prior to analysis. The text of the exam, as well as the code and data used in this analysis, is available in the GitHub repository associated with this paper.

We investigate topics within the text by employing the function \texttt{textmodel\_lda} \cite{ldamodel} in the R package \texttt{seededlda} \cite{seededlda}. Briefly, \texttt{textmodel\_lda} uses Latent Dirichlet Allocation (LDA) to discover groups of words that frequently appear together, thus making it possible for an analyst to determine thematic structures within the text. LDA is a probabilistic model where each document is represented as a mixture of topics, and each topic is represented as a mixture of words. LDA attempts to infer the latent topics and the distribution of those topics within each document.

More formally, let \( K \) be the number of topics, \( M \) be the number of documents in the corpus (16 in each, in our case), and \( N_d \) be the number of words in document \( d \). Define a $k\times V$ matrix $\beta$ where $\beta_{ij} = p(w^j=1\vert z^i = 1)$, where $V$ is the size of the vocabulary, ${\bf w}=(w_1, \ldots, w_N)$ are words within a document, and $z_k$ represents the $k^{th}$ topic. The $v^{th}$ word in a vocabulary is represented by a unit-basis vector such that $w^v = 1$ and $w_u=0$ when $u\neq v$ \cite{blei2003}. Then, \( \theta_d \sim \text{Dirichlet}(\alpha)\) is the topic distribution for document \( d \), \( \phi_k \) is the word distribution for topic \( k \), \( \phi_k \sim \text{Dirichlet}(\beta) \), \( z_{d,n} \) is the topic assignment for the \( n \)-th word in document \( d \), where \( z_{d,n} \sim \text{Multinomial}(\theta_d) \), and \( w_{d,n} \) is the \( n \)-th word in document \( d \), where \( w_{d,n} \sim \text{Multinomial}(\phi_{z_{d,n}}) \). The joint distribution of a topic mixture \( \theta_d \), a set of \( k \) topics \( z_d \), and a set of \( N_d \) words \( w_d \) in a document \( d \) is given by:
\[
p(\theta_d, z_d, w_d | \alpha, \beta) = p(\theta_d | \alpha) \prod_{n=1}^{N_d} p(z_{d,n} | \theta_d) p(w_{d,n} | z_{d,n}, \beta).
\]
\noindent Here, $\alpha$ is a parameter that is a $k$-vector where all components are strictly greater than 0. The parameters $\alpha$ and $\beta$ are ``corpus level parameters'' that are sampled once during corpus creation. Then the marginal likelihood of a document \( d \) is obtained by integrating over \( \theta_d \) and summing over \( z_d \):
\[
p(w_d | \alpha, \beta) = \int p(\theta_d | \alpha) \left( \prod_{n=1}^{N_d} \sum_{z_{d,n}} p(z_{d,n} | \theta_d) p(w_{d,n} | z_{d,n}, \beta) \right) d\theta_d , 
\]
\noindent and the likelihood of the entire corpus \( D \) (with \( M \) documents) is then the product of the marginal likelihoods of the individual documents:
\[
p(D | \alpha, \beta) = \prod_{d=1}^{M} p(w_d | \alpha, \beta)
\]
In practice, estimation of \( \theta \), \( \phi \), and \( z \) is done using methods like Variational Inference or Gibbs Sampling because the exact posterior is intractable \cite{ldamodel}.






\section{Results}\label{sec:results}
The mean score on the graduate exam without including scores for GPT3.5, GPT4, and GPT4o-mini is 69.4, the median is 72.0, and the first and third quartiles are 58 and 87, respectively. The score for GPT3.5 is 41, the score for GPT4o-mini is 72, and the score for GPT4o is 82. GPT3.5 clearly failed the exam, while GPT4 has a score well above the mean. GPT4o-mini performs almost as well as GPT4. GPT3.5 cannot read images or tables, and GPT4o-mini had a somewhat difficult time with images, also, as images cannot be uploaded into GPT4o-mini; however, GPT4o-mini seems to understand the concept of an image (whether a table or a plot) and can give an example with fake numbers. None of the images had alt-text within them; therefore, this is not an explanation for the ability of GPT4o-mini to handle images. For questions using tables and charts, GPT3.5 and GPT4o-mini gave hints for reading the visualizations. For example, when asked to compare the medians of two boxplots, both platforms explained where to find the median of a boxplot in general terms. Generally, GPT4o-mini's explanations were more accurate and nuanced than those of GPT3.5. For example, when asked whether the mean of a data set depicted in a boxplot was less than the median, GPT4o-mini understood that the relationship of the mean and median is dependent on the skewness of the data set. GPT3.5 metaphorically threw up its hands and said that there was no way to determine the relationship between the mean and median from a boxplot because the mean is not plotted on a boxplot.  

\subsection{Results on Descriptive Text Analytics}
In addition to examining the accuracy of the answers to the exams from the three generative AI platforms, we also examined basic text analytics, such as reading level and word frequency. Both reading level and word frequency have bearing on the ease of understanding of the output. Ease of understanding is important because, if the response is written in a fashion that is difficult to understand, then it will not be helpful as a tutor for statistical methods. 

One way to determine legibility is to calculate the grade-level of the answers. There a many methods of calculating grade level; two of the more common are the Flesch-Kincaid \cite{fk} index (FK) and ``Simple Measure of Gobbledygook'' (SMOG)  \cite{smog} indices. The formula for the Flesch-Kincaid measure is $$ FK = 0.39\frac{w}{s} + 11.8\frac{y}{w} - 15.59, $$ where w = total words, s = total sentences, and y = total syllables \cite{fk}. SMOG is calculated using $3\sqrt{y^{\prime}}$ where $y^{\prime}$ is the number of words with multiple syllables within the text \cite{smog}. Both measures align with the U.S. grade level education that readers would need to understand the text. For example, a FK level of 10.2 indicates that the reader would need at least a 10th grade education in the US to understand the text. Generally, FK and SMOG indices greater than 12 indicate the need for a college education, and indices greater than 16 indicate the need for a graduate education. 

Both FK and SMOG measures were calculated for the answers to each question from each platform using \cite{charactercalculator}. Paired t-tools were used to calculate  95\% confidence intervals for the population mean for each platform and reading level index combination. Table \ref{tab:cis} shows the results for FK and SMOG reading levels for GPT3.5, GPT4o-mini, and GPT4. All of the confidence intervals overlap, indicating that it is plausible that the reading levels for each platform are equivalent. Note that the reading levels, except for one, are above twelfth grade, and the upper confidence limits often indicate a graduate level education is necessary to understand the text. While this is somewhat disconcerting, it is possible to ask ChatGPT to respond at a certain grade level. We did not engineer the prompts to do so; we simply took the unedited response to each question without asking for elaboration or clarification. Indeed, the reading level of the prompt can affect the reading level of the response \cite{amin2024}. The issue of the association of the reading level of the prompt with that of the output will be discussed further in Section \ref{sec:disc}.
\begin{table}
  \caption{Descriptive statistics for the generative AI platforms. Given are 95\% t-based confidence intervals for the population mean reading level for both FK and SMOG indices, the mean number of tokens, and the mean number of sentences.}%
  \label{tab:cis}
  \centering
\begin{tabular}{lcccc}
\toprule
Platform & CI-FK & CI-SMOG & Tokens & Sentences \\
\midrule
GPT3.5& (12.8, 15.1) & (14.6, 16.4) & 153 & 7.06 \\
GPT4 & (12.1, 15.4) & (13.5, 16.1) & 99.6 & 4.31\\
GPT4o-mini & (9.3, 22.0) & (14.1, 17.4) & 593 & 20.8 \\
\bottomrule
\end{tabular}
\end{table}

Table \ref{tab:cis} includes the mean number of tokens and sentences for each of the platforms. To obtain these measures, unedited text from each of the 16 questions on the graduate exam was imported into a corpus in R \cite{rcore} using the \texttt{quanteda} package \cite{quanteda}. After removing stopwords, symbols, numbers, punctuation, and single tokens, the number of tokens and number of sentences were calculated for each platform and question separately and aggregated into a mean value for each platform. Single tokens were removed due to the high frequency of tokens such as ``n'' and ''x'', often used as variables to indicate the sample size and an observation, respectively.

Interestingly, GPT4o-mini gives the most complex answers of the three platforms in terms of reading level, number of tokens, and number of sentences. GPT4 tends to be the least complex. Answers from short-answer questions using GPT4o-mini often included background theory and an example before getting to the actual answer of the question. Multiple choice questions were answered succinctly using GPT4o-mini were answered similarly to multiple-choice questions from the other two platforms: the letter of the answer was given with a brief rationalization for the choice.
\begin{figure}[htbp]
\centering
\includegraphics[scale=.2]{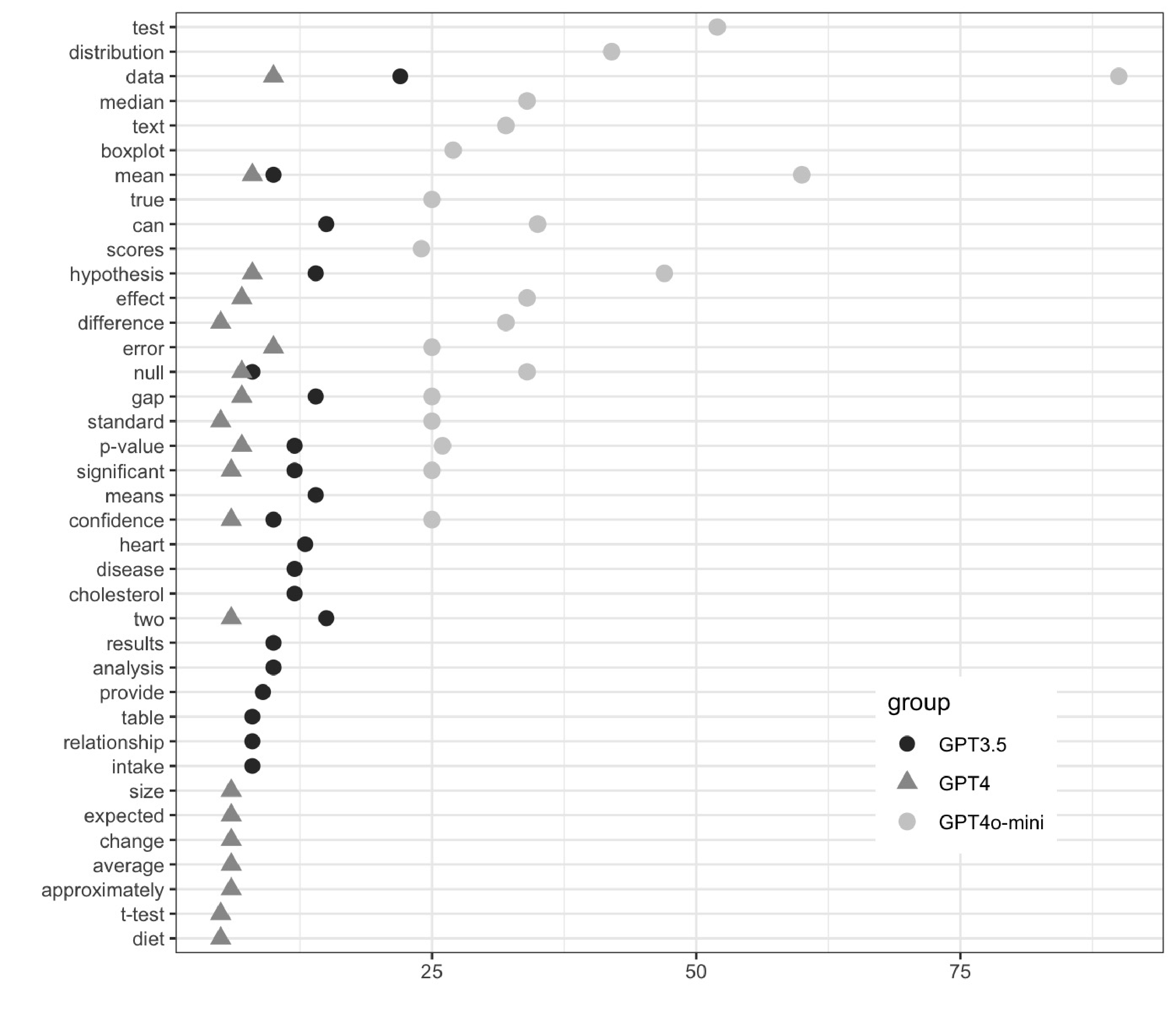}
\caption{Word Frequency comparison of GPT3.5 (black circles), GPT4 (dark gray triangles), and GPT4o-min (light gray circles).}\label{fig:ffplot}
\end{figure}

Figure \ref{fig:ffplot} shows the top 20 most frequently used words extracted from the response to the short-answer and discussion questions for GPT3.5, GPT4, and GPT4o-mini. Black circles indicate frequencies from GPT3.5, medium gray triangles indicate frequencies from GPT4, and light grey circles indicate  frequencies from GPT4o-mini. Generally, all three platforms share their highest frequency words in approximately the same frequency order, and most of the words are words that one would expect to be common in a statistics course. Also generally, ChatGPT4o-mini tends to use the same words more frequently than do either GPT3.5 or GPT4. This is due to its tendency to explain how to work a question in addition to providing the answer. 

Some exceptions to the use of statistical terms are the words ``heart'', ``disease'', and ``cholesterol''. For GPT4o-mini or GPT4, these words do not appear in the top 20 frequently used words. However, each of these words are used multiple times by GPT3.5. This usage is due to the fact that one of the problems involved interpretation of a $\chi^2$ test for the association between heart disease and cholesterol. The increase in frequency for GPT3.5 can be attributed to the fact that the data were given in a table that it could not read. Instead of performing a $\chi^2$ test of independence as requested by the problem, GPT3.5 apologized for being unable to read the table, and proceeded to explain about the evidence from the literature for an association between heart disease and cholesterol. In short, it behaved as if it were a student who did not know how to answer the question, but wanted to gain some points by talking about the known relationship between heart disease and cholesterol.
\begin{figure}[htbp]
\centering
\includegraphics[scale=.4]{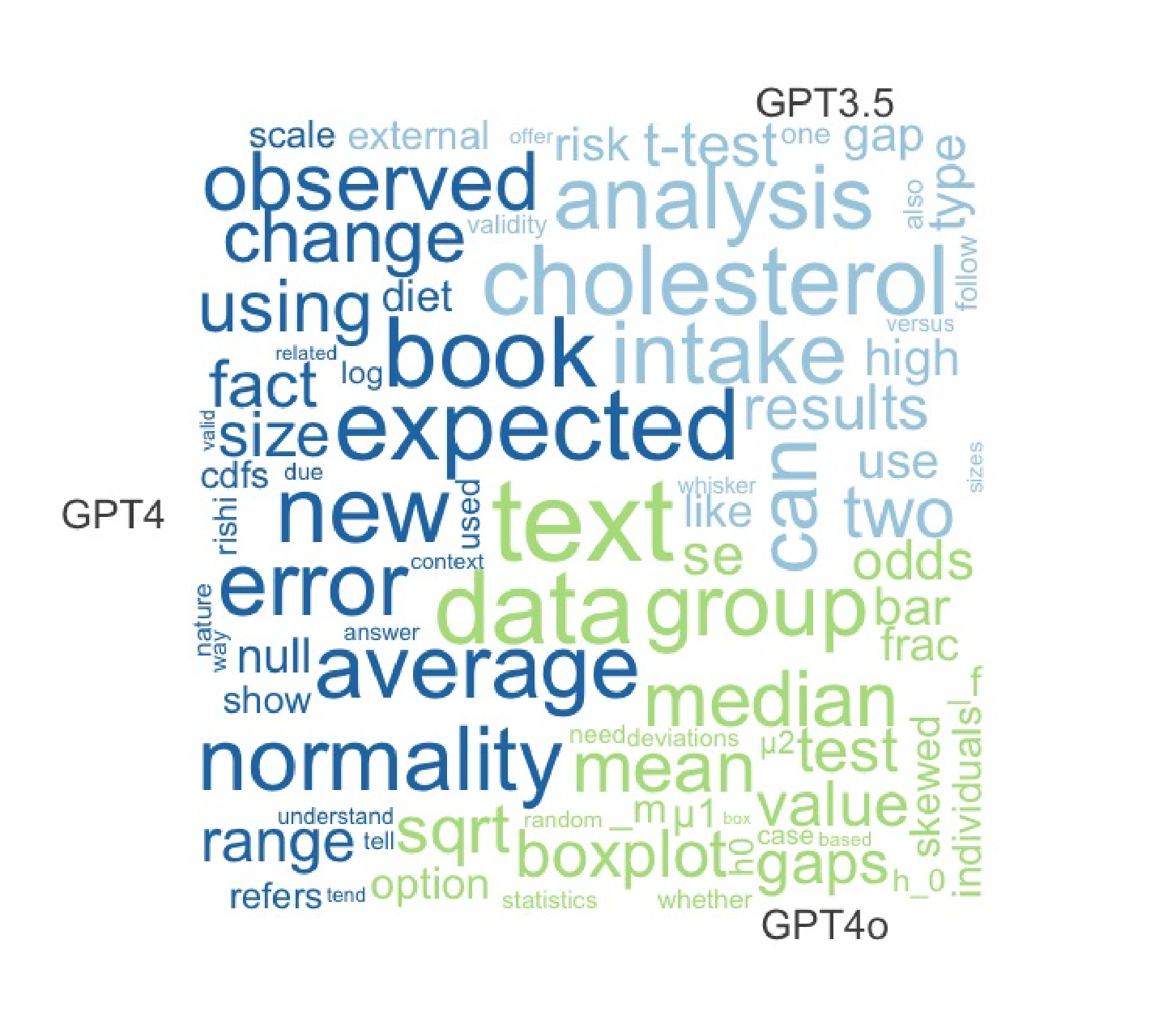}
\caption{Comparison Cloud of frequency of words from GPT3.5 (light blue, top right of cloud), GPT4 (dark blue, left side of cloud), and GPT4o-mini (light green, bottom right of cloud).}\label{fig:cloud}
\end{figure}

Word frequency data is displayed another way in Figure \ref{fig:cloud}. Here, we limited the display to words that appeared a minimum of 10 times. The size of the font for a given word is proportional to how often it appears within its platform group. With the word cloud, the differences among the frequencies of words among the generative AI platforms is clearer. The word ``cholesterol'', for example, jumps out in the middle of the plot for GPT3.5, but does not appear for the other platforms. In this representation, it seems that collection of most frequent words for each platform is quite different. We investigate the differences further with topic modeling. 

\subsection{Results on Topic Modeling}
Topic modeling is a statistical method used to uncover hidden themes (topics) within a large collection of documents. It identifies groups of words that frequently appear together, allowing for the discovery of underlying topics without prior labeling, useful for summarizing and understanding large text datasets. For our application, topic modeling helps us determine whether the responses to the questions are relevant to the topic of statistical methods. If topics are unrelated to statistical methods, or unrelated to the questions being asked, then we can conclude that ChatGPT might be at worst, hallucinating, or, at the least, providing responses that are irrelevant to the question being asked. The seeds of topic modeling were planted mid-way through the last century \cite{ballester2022}. Since that time, other methods such as non-negative matrix factorization (NMF) \cite{leeseung}, latent Dirichlet allocation (LDA) \cite{blei2003}, and Word2Vec \cite{ballester2022} have been developed. For our investigation, we have chosen LDA due to its relative simplicity and the ready availability of R code. 

For our application, a ``corpus'' refers to the text answers to the graduate exam questions for one version of the generative AI platforms examined. Each corpus in our application contains 16 documents. We perform LDA on each corpus separately and examine the results to determine whether the same topics are discovered within each corpus. Because we are unsure about the number of topics in our corpus, we use unsupervised LDA with $k=5$ as a starting point. $k=5$ was used because each corpus is small, consisting of only 16 documents. The choice of $k$ has been investigated in several papers, for example, \cite{ballester2022}, \cite{meaney2019}, and \cite{hecking2018}. Most investigations of the choice of $k$ on the robustness of LDA have used corpora with hundreds or thousands of documents \cite{hecking2018}. In general, the larger the value of $k$, the more granular the results, and the more likely that nonsensical topics will be ``discovered'' by the algorithm. We began with  $k = 5$, but discovered that our documents tended to map to only 3 topics; therefore, we settled with $k=3$. A small value of $k$ is appropriate for small corpora. Table \ref{tab:topics} shows the topics and their associated terms for the three generative AI models.

\begin{table}[htbp]
\centering
    \caption{List of topics and associated terms for all three generative AI models for LDA with $k=3$. Topics are not named because each topic is different for each generative AI verison.}\label{tab:topics}
    \begin{tabular}{clll}
\toprule
& topic1 & topic2 & topic3\\
\midrule
&heart & mean & p-value\\
&cholesterol & values & hypothesis\\
&disease & confidence & two\\
&data & data & distribution\\
GPT3.5 &analysis & results & gap\\
&can & can & test\\
&relationship & statistical & standard\\
&intake & effect & error\\
&boxplots & intervals & null\\
&information & provide & type\\
\midrule\midrule
&average & mean & hypothesis\\
&two & effect & null\\
&test & data & approximately\\
&sample & confidence & p-value\\
GPT4 &size & scores & gap\\
&error & range & book\\
&using & values & significant\\
&different & results & observed\\
&textbooks & significance & change\\
&distribution & log & difference\\
\midrule\midrule
& mean & effect & data\\
& median & hypothesis & test\\
& data & p-value & distribution\\
& boxplot & null & text\\
GPT4o-mini & can & confidence & standard\\
& values & data & gap\\
& original & interval & difference\\
& outliers & p-values & diet\\
& observational & type & mean\\
& studies & test & group\\
\bottomrule
\end{tabular}
\end{table}
From our topic modeling analysis, we see that the three topics are not consistent for the three generative AI versions. Topic 1 for GPT3.5 seems to be more about the discipline-specific application of the problems in the corpus, while topic 1 for GPT4 and GPT4o-mini are about descriptive statistics. This somewhat matches word frequency findings displayed in Figures \ref{fig:ffplot} and \ref{fig:cloud}, in that the more frequent words from GPT3.5 had to do with the context of the problem rather than statistical methodology. Topic 2 for GPT3.5 has to do with descriptive statistics, although words from inferential statistics, such as ``confidence''  are in the list. Topic 2 for GPT4 concerns data analysis and Topic 3 concerns more inferential statistics. Topics 2 and 3 for GPT4o-mini are more like those of GPT4 than they are of GPT3.5. There is only one application word, ``textbooks'' that appears in the topic lists for GPT4 and GPT4o-mini. While this investigation involves very small corpora, there is evidence that GPT4 and GPT4o-mini are more alike than either of them is to GPT3.5. This finding could be construed as more sophistication in statistical analysis with the evolution of the ChatGPT platform; however, more investigation on larger corpora is needed.

\section{Discussion and Conclusion}\label{sec:disc}

This work is not advocating that professors allow students to use generative AI on exams. Rather, an exam serves as a convenient way to measure differences in accuracy and characteristics of text output in free and paid versions of a popular generative AI platforms. GPT3.5, now deprecated, has been replaced by GPT4o-mini, and some of the inaccuracies remain. GPT4 performs much better in terms of accuracy. If the results are similar with replication on multiple exams in statistics courses throughout a typical statistics and data science curriculum, then administrators and educators need to examine ways to make better performing versions accessible to disadvantaged students. Equity in the use of AI might become less of an issue in the future as current subscription versions are replaced by more powerful upgrades and the older versions become free for use. 

This study's analysis of topic modeling in the responses generated by GPT3.5, GPT4, and GPT4o-mini reveals distinct differences in how these AI models describe statistical concepts. Topic modeling indicated that, while all three versions exhibited distinct differences, GPT4 and GPT4o-mini produce more thematically consistent and statistically focused content compared to GPT3.5, which also often provided less accurate responses.  GPT4 tended to focus more on data analysis and inferences while GPT3.5 tended to focus mostly on descriptive statistics.  This suggests that the newer models, particularly GPT4, have improved capabilities in generating more relevant and complete responses to questions from this statistics exam. 

This comparison of accuracy and topic modeling among GPT3.5, GPT4, and GPT4o-mini is limited in a number of ways. First, only one exam with 16 questions was used for comparison, and the exam was geared toward first-year PhD-level students. A future comparison of multiple similar exams is needed to determine whether the results hold in a larger population of exams, and the results could be extended to different levels of exams. For example, this particular exam was focused on methods and did not contain equations or theory. Therefore, the results do not apply to more theoretical statistical courses. 

Recent research has revealed that small changes to prompts can change the response in generative AI platforms \cite{joshi, callanan2023, ball2024can, yao2024}. Indeed, it is unlikely that students using ChatGPT as a tutor will copy and paste homework questions will use a ``zero-shot'' framework \cite{yao2024}, as we have done here. Students will probably ask ChatGPT to explain further, or to simplify the answer in some fashion. However, it is unlikely that students will think to provide context such as, ``I am a student in a first-year statistics course'' \cite{joshi}, when they are using ChatGPT for help. Educators can help students who use generative AI outside of the classroom by informing them that providing context helps, and ChatGPT can be asked to clarify or simplify results. However, it is possible that asking for a simplification can read to inaccuracy and further misunderstanding.  As an illustration, we  asked GPT4o (the latest version) to explain the concept of statistical significance in language that a fifth grader could understand. Here is the response:

\begin{quote}
Alright,  imagine you're in a candy shop. You want to know if people like chocolate or gummy bears better. You ask ten of your friends, and seven say chocolate while three say gummy bears. But here's the thing: What if this group of ten friends doesn't really represent everyone else? Maybe it's just a coincidence that most of them like chocolate.

Statistical significance is like a test to figure out if what you're seeing is really true for most people or if it just happened by chance. Scientists use math to check this. If the math says, "Yep, this is pretty solid," then we say the result is statistically significant. It's like saying, "This isn't just a lucky guess. It's probably true for a bigger group."

If the math says, "Hmm, it could just be luck," then we say it's not significant, and we can't be sure if chocolate is really more popular than gummy bears.
\end{quote}

ChatGPT's answer is certainly helpful, but it is not quite as precise as a college professor would like. Next, we prompted GPT4o with ``Same question, but for a first-year college student.''. Here is the response,

\begin{quote}
Statistical significance is a way to determine whether the results of a study are likely to reflect something real or are just due to random chance. Let's say you conduct a survey asking students if they prefer coffee or tea, and 70\% say coffee. That sounds like most people prefer coffee, but you need to ask: could this result just be random, based on the specific group of people you asked?

To test this, we use statistical tools to calculate a p-value. The p-value tells us the likelihood of seeing our result (or something more extreme) if there's really no difference between preferences for coffee and tea in the population. If the p-value is small (typically less than 0.05), we say the result is statistically significant. This means it's unlikely the result happened by chance, and we have stronger evidence that most students do prefer coffee.

If the p-value is large, we can't confidently say there's a real difference. It could just be luck. Statistical significance doesn't mean the result is important or practical; it just means it's unlikely to be random.
\end{quote}

The responses regarding statistical significance are not completely accurate. There are many papers that caution against the use of the phrase ``statistically significant'' in the literature and against the use of a p-value cutoff for the declaration of statistical significance \cite{wass2016}. Certainly the effect of reading level of the prompt on accuracy of the response is another issue to be addressed in future work.

Interestingly, the Flesch-Kincaid reading level for the response where ChatGPT was asked to respond at a fifth-grade level is 6.5 and the reading level for the first year college student response is 10.4. Therefore, it appears that the reading level of the response is somewhat associated with the reading level of the prompt.  \cite{amin2024} have investigated the use of GPT3.5, GPT4, Bard, and Bing in their ability to provide easier to read health information for pediatric populations. They found that LLMs can tailor their responses to different reading levels, but have difficulty producing outputs below a sixth-grade level, even when asked to do so. The association of reading level of the prompt with the reading level of the response has bearing on this work, as the questions entered into ChatGPT were at a graduate school level. It seems reasonable that ChatGPT would respond at the same, or at a similar level. 

Another obvious extension is to perform this experiment using other generative AI platforms, both paid and free. GPT4o \cite{gpt4o} and future updates, are obvious candidates, as well as Google Gemini \cite{gemini} and Anthropic Claude \cite{claude}, in both their free and subscription versions. Microsoft Co-Pilot \cite{copilot} is available for those who have access to Office365. In addition, there are open source platforms, mostly hosted on HuggingFace \cite{wolf2020huggingfaces}, such as LLaMA-2 \cite{meta, touvron2023}, Colossal AI \cite{colossalAI},  and OpenChatKit \cite{openchatkit}. Because the open source AI platforms are customizeable, it would be impossible to examine all iterations, but they are freely available, which would potentially allow for more fine-tuned tutoring for statistical methodology. 

In addition, there is room for methodological development in text analytics. When using LDA for topic modeling, the resulting topics tended to change slightly as the algorithm was rerun. Therefore, there is variability in the results of topic modeling, and it would be useful to quantify this in some fashion and use it to build inferential statistics. Other extensions include use of methods for topic modeling other than LDA, to determine the sensitivity and robustness of results to various methods. Further, if multiple tests are used, the questions will be nested within the test, and methods of topic modeling for documents nested within corpora will need to be developed.

Finally, we began this paper with issues around concerns about the promise of equitable education using generative AI platforms because of lack of access to paid platforms, and a possible difference in performance of free-tier vs paid-tier platforms. We have shown a clear difference in performance for GPT3.5, GPT4o-mini (which are free), and GPT4 (costing \$20 per month). How do our results inform this conversation? Educators could ask their students to refrain from using paid services or from using generative AI at all; however, this is hard to police and counterproductive. At the institutional level, access to paid AI tools could be made freely available for all students; however, the financial commitment could be substantial. As of the writing of this manuscript, OpenAI does not have education-level pricing, although Microsoft Co-Pilot is available for institutions that subscribe to Office365. For an institution of 10,000 students to make GPT4o (the latest version) free for all students, the institution would have to pay \$20/month for each student or \$200,000 per month, which is not sustainable. Therefore, there is an opportunity for institutions to talk to AI providers about educational pricing. Other possibilities are having students use AI only in the classroom, in a flipped classroom model, where educators can ensure students are using AI correctly, modifying prompts and necessary, and understanding the responses. This would also help educators know whether the responses are correct, and educators can clarify or correct AI responses in real time. Another model would be for educators and students to think of a subscription to a generative AI platform as we currently think of textbook rentals, which might be covered by financial aid. While the world of generative AI opens many possibilities for closing digital and educational gaps, many issues, including the more accurate performance of paid versions of generative AI, need to be resolved before the promise of AI is realized.

\bibliographystyle{plain}
\bibliography{JDSbib}

\end{document}